\documentstyle[aps,prd]{revtex}

\def\be{\begin{equation}}
\def\ee{\end{equation}}
\def\bea{\begin{eqnarray}}
\def\eea{\end{eqnarray}}

\begin{document}
\draft
\title{GEOMETRIC CONDITIONS ON THE TYPE OF MATTER DETERMINING \\
THE FLAT BEHAVIOR OF THE ROTATIONAL CURVES IN GALAXIES }
\author{Tonatiuh Matos$^a$, Dar{\'\i}o N{\'u}{\~n}ez$^b$, F. Siddhartha Guzm{\'a}n$^a
$, Erandy Ram{\'\i}rez$^b$}
\address{$^a$Departamento de F{\'\i}sica, \\
Centro de Investigaci{\'o}n y de Estudios Avanzados del I. P. N.,\\
A. P. 14-700, 07000 M{\'e}xico, D.F.,MEXICO\\
$^b$Instituto de Ciencias Nucleares, \\
Universidad Nacional Aut{\'o}noma de M{\'e}xico\\
A. P. 70-543, 04510 M{\'e}xico, D. F., MEXICO}
\date{\today}
\maketitle

\begin{abstract}
In an arbitrary axisymmetric stationary spacetime, we determine the
expression for the tangential velocity of test objects following a circular
stable geodesic motion in the equatorial plane, as function of the metric
coefficients. Next, we impose the condition, observed in large samples of
disks galaxies, that the magnitude of such tangential velocity be radii
independent, obtaining a constraint equation among the metric coefficients,
and thus arriving to an iff condition: The tangential velocity of test
particles is radii independent iff the metric coefficients satisfied the
mentioned constraint equation. Furthermore, for the static case, the
constraint equation can be easily integrated, leaving the spacetime at the
equatorial plane essentially with only one independent metric coefficient.
With the geometry thus fixed, we compute the Einstein tensor and equate it
to and arbitrary stress energy tensor, in order to determined the type of
energy-matter which could produce such a geometry. Within an approximation,
we deduce a constraint equation among the components of the stress energy
tensor. We test in that constraint equation several well known types of
matter, which have been proposed as dark matter candidates and are able to
point for possible right ones. Finally, we also present the spherically
symmetric static case and apply the mentioned procedure to perfect fluid
stress energy tensor, recovering the Newtonian result as well as the one
obtained in the axisymmetric case. We also present arguments on the need to
use GR to study types of matter different than the dust one. 
\end{abstract}


\section{Introduction}

One of the most important achievements of the present cosmology is doubtless
the complete identifications and the accounting of the amounts of the
different types of matter and energy which are present in the Universe (for
an excellent review see for example \cite{turner1}). Essentially, the
present components of the Universe are composed by matter and vacuum energy $%
\Omega _{0}=\Omega _{M}+\Omega _{\Lambda }$ \cite{pee1}. Furthermore, there
is a very good evidence that the Universe is flat. That evidence comes
either from the theory, where the most accepted model for the early Universe
is inflation, as well as from observational data, which implies that $\Omega
_{0}=1\pm 0.12 $ (see for example \cite{sonido}). The mass of the galaxy
clusters is perhaps the most reliable way for determining the matter
component $\Omega _{M}$. Observations indicate that $\Omega _{M}\sim 0.3$ 
\cite{turner1}, however, the main visible components of $\Omega _{M}$,
baryons, neutrinos, form a very small fraction of $\Omega _{M}$.
Observations indicate that stars and dust (baryons) represents something
like $5\%$ of \ the whole matter of the Universe. In other words, $\Omega
_{M}\sim \Omega _{b}+\Omega _{DM}\sim 0.05+\Omega _{DM}$, where $\Omega _{DM}
$ represents the dark matter part of the matter contributions which has to
have a value of $\Omega _{DM}\sim 0.25$. Notice that the given value of the
amount of baryonic matter is in concordance with the limits imposed by
nucleosynthesis (see for example \cite{schram}). A greater amount of
baryonic matter could change the predicted values of primordial H and $^{4}$%
He in the standard model of cosmology which coincide very well with
observations.\newline

The existence of dark matter in the Universe has been firmly established by
astronomical observations at very different length-scales, ranging from the
local galaxies to clusters of galaxies. The standard way to notice this need
for dark matter comes within the framework of mechanics: A large fraction of
the mass needed to produce the observed dynamical effects in all these very
different systems, is not seen. This puzzle has stimulated the exploration
of several proposals, and very imaginative explanations have been put
forward, from exotic matter like supersymmetric particles \cite{susy} to non
relativistic modifications of Newtonian dynamics \cite{mond} and non-linear
general relativistic theories \cite{kazanas}. Above this, it is believed
that this dark matter is such that interacts very weakly with ordinary
matter, which makes it very difficult to detect by other means other than by
their gravitational effects on the baryonic matter as it is well stablished
for the cold dark matter scenario \cite{nfw,nfw2}. The bottom line is that
one of the most important components of the density of the Universe, the
dark matter, still eludes us, and the question remains open: Which is the
nature of the dark matter component?\newline

At the galactic scale, the problem is clearly posed: The measurements of
rotation curves (RC) in galaxies show that the coplanar orbital motion of
gas in the outer parts of these galaxies keeps a more or less constant
velocity up to several luminous radii \cite{persic,persic2}. The discrepancy
arises when one applies the usual Newtonian dynamics to the observed
luminous matter and gas, since then, the circular velocity should decrease
as we move outwards. The most widely accepted explanation is that there
exists a spherical halo of dark matter, its nature being unknown, which
surrounds the galaxy and account for the missing mass needed to produce the
flat behavior of the RC.\newline

The main goal of this work is to study the dark matter problem in spiral
galaxies, using a fully relativistic approach, continuing with the work
started by Matos and Guzm{\'{a}}n \cite{tofc}, where they made a preliminary
dynamical analysis in the context of spiral galaxies. We apply a deductive
method: Starting from a reasonable general space-time, we deduce, in terms
of arbitrary metric coefficients the expression for the tangential velocity
of test objects following a circular stable geodesic motion in the
equatorial plane. We impose next the condition, observed in hundreds of
galaxies, that such tangential velocity be radii independent, and obtain a
constraint equation among the metric coefficients, Eq.(\ref{eq:con1}).
Arriving in this way to an iff condition: In an axisymmetric static
space-time, the tangential velocity of test particles at the equatorial
plane is radii independent iff the metric coefficients satisfied Eq.(\ref
{eq:con1}). Furthermore, for the static case, the constraint equation can be
solved, leaving the space-time essentially with only one independent metric
coefficient, thus determining very narrowly the type of space-time which can
have a geometry such that the tangential velocity of the geodesics of test
particles moving in stable circular equatorial orbits, be radii independent.
With the geometry thus fixed, we compute the Einstein tensor and equate it
to and arbitrary stress energy tensor, in order to determined the type of
energy-matter which could produce such a geometry. We are able to deduce a
constraint equation among the components of the stress energy tensor, Eq.(%
\ref{eq:fin}). Using that constraint equation, Eq.(\ref{eq:fin}), we test
several well known types of matter, which have been proposed as dark matter
candidates and are able to impose further restrictions on most of them as
possible candidates for the dark matter in the region where it is observed
the mentioned behavior on the tangential velocity. It is stimulating that
the gravitational physics has been developed to such degree in which we can
actually follow Sherlock Holmes' maxim: ``...when you have excluded the
impossible, whatever remains, however improbable, must be the truth'' \cite
{Sher}.\newline

The work is composed as follows: In section 2 we determine the geometry of
axisymmetric static space times allowing a tangential velocity with a
magnitude independent of the radius. In section 3, we work with the Einstein
equations for the geometry thus determined and for a general stress energy
tensor, obtaining a constraint for the components of such stress energy
tensor, and we test four types of matter into that constraint, being able to
restrict most of them, among which is the perfect fluid. Finally, in the
conclusion we discuss our results and propose a coupling of types of matter,
which could be the one we are looking for in order to determine the nature
of the dark matter. We also present this analysis for the spherical static
space time in appendix \ref{app:1} and for the axisymmetric stationary case
in appendix \ref{app:2}.\newline

\section{Form of the line element}

In this section we study the conditions which the flatness of the tangential
velocity of the RC, imposes on the metric coefficients. We want to stress
the fact that the results presented in this section are independent of type
of energy-matter tensor present in the space-time and curving it. It is a
purely geometric analysis.\newline

As mentioned above, observational data show that the galaxies are composed
by almost 90\% of dark matter. Thus we can suppose that luminous matter does
not contribute in a determining way to the total energy density of the
galaxy, at least in the region where the flatness of the RC is observed.
Consequently, we consider that the dark matter will be the main contributor
to the dynamics, and we will treat the observed luminous matter as a test
fluid, that's it, in this approximation we will neglect the contribution of
the luminous matter to the curvature, {\it i. e.}, to the dynamics. Also, it
is reasonable to suppose that the halo of dark matter is symmetric with
respect to the rotation axis of the galaxy, thus we take the symmetry of the
space-time as axially symmetric. Furthermore, the observations allow us to
take the space-time as stationary as well. Thus, the most general reasonable
space-time which we can study is an axisymmetric stationary one. The
line-element of such space-time, given in the Papapetrou form is: 
\begin{equation}
ds^{2}=-e^{2\,\psi}(dt+\omega \,d\varphi)^{2}+e^{-2\psi }\,
[e^{2\gamma}(d\rho^{2}+dz^{2})+\mu^{2}d\varphi^{2}],  \label{eq:ele}
\end{equation}
where $\psi ,\omega ,\gamma $, and $\mu $, are functions of $(\rho ,z)$.%
\newline

We will derive the geodesic equations in the equatorial plane, that is for $%
z=\dot{z}=0$, where dot stands for the derivative with respect to the proper
time, $\tau $. Then, we will study the constrains imposed for circular
geodesics on the energy and angular momentum of particles in such orbits,
and obtain an expression for the tangential velocity for the particles
moving along those geodesics, described in terms of the metric coefficients.
Finally, we impose the condition that such tangential velocity be radii
independent and derive the restriction that then has to be satisfied among
the gravitational coefficients. In this section we derive the expressions
for a static space time and in the Appendix \ref{app:2} we do so for the
stationary one. We split the presentation in this way not only for the sake
of clarity, but from the fact that the static case is also quite realistic,
considering that the observed velocity of the stars orbiting a galaxy in the
region of interest is quite non relativistic (of the order of 230 Km per s),
thus we can infer that the space-time is not very rapidly rotating.\newline

The Lagrangian for a test particle traveling on the static space time ($%
\omega =0$) described by (\ref{eq:ele}) is given by: 
\begin{equation}
2{\cal {L}}=-e^{2\,\psi }\dot{t}^{2}+e^{-2\psi }[e^{2\gamma }\,(\dot{\rho}%
^{2}+\dot{z}^{2})+\mu ^{2}\,\dot{\varphi}^{2}],  \label{eq:lag}
\end{equation}
thus, the associated canonical momenta, $p_{x^{a}}={\frac{{\partial {\cal {L}%
}}}{{\partial \dot{x^{a}}}}}$, are: 
\begin{eqnarray}
p_{t}=-E &=&-e^{2\,\psi }\dot{t}, \\
p_{\varphi }=L &=&\mu ^{2}\,e^{-2\,\psi }\dot{\varphi}, \\
p_{\rho } &=&e^{-2(\psi -\gamma )}\dot{\rho}, \\
p_{z} &=&e^{-2(\psi -\gamma )}\dot{z},  \label{eq:mta}
\end{eqnarray}
where $E$, and $L$, are constants of motion for each geodesic, a fact which
comes from the symmetries of the space-time analyzed. As there is no
explicit dependence on time, $t$, the Hamiltonian, ${\cal {H}}=p_{a}\,\dot{%
x^{a}}-{\cal {L}}$, is another conserved quantity, which we normalized to be
equal to minus one half for time-like geodesics. Also, we restrict the
motion to be at the equatorial plane, thus $\dot{z}=0$. In this way, we
obtain the following equation for the radial geodesic motion: 
\begin{equation}
\dot{\rho}^{2}-e^{2(\psi -\gamma )}\,[E\,\dot{t}-L\,\dot{\varphi}-1]=0.
\label{eq:ham}
\end{equation}

In order to have stable circular motion, which is the motion we are
interested in, we have to satisfy three conditions:

i) $\dot{\rho}=0$, and

ii)${\frac{{\partial V(\rho)}}{{\partial\,\rho}}}=0$, where $%
V(\rho)=-e^{2(\psi-\gamma)}\,[E\,\dot{t}-L\,\dot{\varphi}-1] $.

iii)${\frac{{\partial^2 V(\rho)}}{{\partial\,\rho^2}}}|_{extr}>0$, in order
to have a minimum.\newline

With these conditions, from Eq.(\ref{eq:ham}), we obtain a set of two
equations constraining the motion to be circular extrema in the equatorial
plane:

\begin{eqnarray}
E\,\dot{t}-L\,\dot{\varphi}-1 &=&0,  \nonumber \\
{\frac{{\partial }}{{\partial \,\rho }}}\left( e^{2(\psi -\gamma )}\,[E\,%
\dot{t}-L\,\dot{\varphi}-1]\right) &=&0.  \label{eq:cons0}
\end{eqnarray}

>From Eq.(\ref{eq:mta}), we can express $\dot{t}$, and $\dot{\varphi}$ in
terms of $E,L$, and the metric coefficients as

\begin{eqnarray}
\dot{t} &=&{e^{-2\psi }}E, \\
\dot{\varphi} &=&{\frac{{e^{2\psi }}}{{\mu ^{2}}}}\,L.  \label{eq:tp}
\end{eqnarray}

Using these equations in the constraints ones and recalling that $E$ and $L$
are constants for each circular orbit, after some rearranging, we arrive to
the following equations:

\begin{eqnarray}
\mu ^{2}\,e^{-2\psi }\,(1-e^{-2\psi }\,E^{2})+L^{2} &=&0, \\
-(e^{-2\psi })_{\rho }\,E^{2}+\left( {\frac{{e^{2\psi }}}{{\mu ^{2}}}}%
\right) _{\rho }\,L^{2} &=&0,
\end{eqnarray}

\noindent where the subindex stands for derivative with respect to ${\rho }$%
. Solving for $E$ and $L$, we obtain:

\begin{eqnarray}
E &=&e^{\psi }\,\sqrt{{\frac{{{\frac{{\mu _{\rho }}}{{\mu }}}-\psi _{\rho }}%
}{{{\frac{{\mu _{\rho }}}{{\mu }}}-2\,\psi _{\rho }}}}},  \nonumber \\
L &=&\mu \,e^{-\psi }\,\sqrt{{\frac{{\psi _{\rho }}}{{{\frac{{\mu _{\rho }}}{%
{\mu }}}-2\,\psi _{\rho }}}}}.  \label{eq:EL1}
\end{eqnarray}

The second derivative of the potential $V(\rho )$ evaluated at the extreme,
in this case means evaluate at the values of $E$ and $L$ which constraint
the motion to be circular and extrema, is given by:

\begin{equation}
V_{\rho \rho }|_{extr}={\frac{{2\,e^{2(\psi -\gamma )}}}{{{{\frac{{\mu
_{\rho }}}{{\mu }}}-2\,\psi _{\rho }}}}}\left( {\frac{{\mu _{\rho }}}{{\mu }}%
}\,\psi _{\rho \rho }-{\frac{{\mu _{\rho \rho }}}{{\mu }}}\,\psi _{\rho }+4\,%
{\psi _{\rho }}^{3}-6{\frac{{\mu _{\rho }}}{{\mu }}}\,{\psi _{\rho }}%
^{2}+3\,\left( {\frac{{\mu _{\rho }}}{{\mu }}}\right) ^{2}\,\psi _{\rho
}\right) .  \label{eq:potg}
\end{equation}

We can now obtain an expression for the angular velocity of a test particle, 
$\Omega $, moving in a circular motion in the orbital plane, in terms of the
metric coefficients, recalling that 
\begin{equation}
\Omega ={\frac{{d\varphi }}{{dt}}}={\frac{{\dot{\varphi}}}{{\dot{t}}}},
\label{eq:om1}
\end{equation}

\noindent thus, using Eqs.(\ref{eq:tp}), and (\ref{eq:EL1}), in this last
equation for the angular velocity, we obtain that:

\begin{equation}
\Omega ={\frac{{e^{2\psi }}}{{\mu }}}\,\sqrt{{\frac{{\psi _{\rho }}}{{{\frac{%
{\mu _{\rho }}}{{\mu }}}-\psi _{\rho }}}}}.  \label{eq:om2}
\end{equation}

Finally, in order to express the tangential velocity of the test particles
in circular motion in the equatorial plane, in terms of the metric
coefficients, following Chandrasekhar \cite{Chan}, we rewrite the line
element given in Eq.(\ref{eq:ele}) as:

\begin{equation}
ds^{2}=-{e^{2\,\psi }}dt^{2}+e^{-2\psi }\,\mu ^{2}\,\,d\varphi
^{2}+e^{-2(\psi -\gamma )}(d\rho ^{2}+dz^{2}),  \label{eq:ele1}
\end{equation}

\noindent thus, in terms of the proper time, $d\tau ^{2}=-ds^{2}$, we have
that

\begin{eqnarray}
d\tau ^{2} &=&{e^{2\,\psi }}\,dt^{2}\,[1-{{\,}e^{-4\,\psi }\,\mu ^{2}}\left( 
{\frac{{d\varphi }}{{dt}}}\right) ^{2}+  \nonumber \\
&&-{e^{2\,\gamma }\,e^{-4\,\psi }}\left( \left( {\frac{{d\rho }}{{dt}}}%
\right) ^{2}+\left( {\frac{{dz}}{{dt}}}\right) ^{2}\right) ],
\end{eqnarray}

\noindent from which we can write that

\begin{equation}
1={e^{2\,\psi }}\,{u^{0}}^{2}\,[1-v^{2}],
\end{equation}

\noindent where $u^{0}={\frac{{dt}}{{d\tau }}}$ is the usual time component
of the four velocity, and a definition of the spatial velocity, $v^{2}$,
comes out naturally in this way.

\begin{equation}
v^{2}=e^{-4\,\psi }\,\mu^2\,\left( {\frac{{d\varphi }}{{dt}}}\right)^{2}+ {%
e^{2\,\gamma }\,\,e^{-4\,\psi }}\left( \left( {\frac{{d\rho }}{{dt}}}%
\right)^{2}+ \left( {\frac{{dz}}{{dt}}}\right)^{2}\right) .
\end{equation}

This spatial velocity is the 3-velocity of a particle measured with respect
to an orthonormal reference system (see section 52 of \cite{Chan}), thus has
components:

\begin{equation}
v^{2}={v^{(\varphi )}}^{2}+{v^{(\rho )}}^{2}+{v^{(z)}}^{2}.
\end{equation}

>From these last two expressions we obtain for the $\varphi -$component the
spatial velocity:

\begin{equation}
v^{(\varphi )}={e^{-2\,\psi }}\mu \,\,\Omega ,  \label{eq:vphi}
\end{equation}

\noindent and substituting $\Omega $ from Eq.(\ref{eq:om2}), we finally
obtain an expression for the tangential velocity of a test particle in
stable circular motion, in terms of the metric coefficients of the general
line element given by Eq.(\ref{eq:ele}), such tangential velocity has the
form:

\begin{equation}
v^{(\varphi )}=\sqrt{{\frac{{\psi _{\rho }}}{{{\frac{{\mu _{\rho }}}{{\mu }}}%
-\psi _{\rho }}}}}.  \label{eq:vphig}
\end{equation}

It was our goal to obtain this expression for the tangential velocity for a
general axisymmetric static space time, and to be able to describe it in
terms of the metric coefficients alone, because now we can impose conditions
on this tangential velocity, and deduce a constraint equation among the
metric coefficients, which has to be satisfied in order to fulfill the
condition imposed on the velocity. In particular, the tangential velocity
for circular trajectories in each orbit is constant, that is $v^{(\varphi
)}{}_{\rho }=0$, thus $v^{(\varphi )}=v_{c}^{(\varphi)}$, with $%
v_{c}^{(\varphi )}$ a constant, representing the value of the velocity, from
Eq. (\ref{eq:vphig}), we have that:

\begin{equation}
{\frac{{\mu _{\rho }}}{{\mu }}}={\frac{{1+{v_{c}^{(\varphi )}}^{2}}}{{{%
v_{c}^{(\varphi )}}^{2}}}}\,\psi _{\rho }.  \label{eq:con0}
\end{equation}

Finally, with respect to the $z$-motion, considering that at the equatorial
plane not only $\dot{z}=0$, but also that $\ddot{z}=0$, that is that the
forces above and below the plane cancel out, from the geodesic $z$-equation,
using Eq.(\ref{eq:con0}), we obtain that this relation among the metric
coefficients must hold for the derivatives with respect to $z$ as well:

\begin{equation}
{\frac{{\mu _{z}}}{{\mu }}}={\frac{{1+{v_{c}^{(\varphi )}}^{2}}}{{{%
v_{c}^{(\varphi )}}^{2}}}}\,\psi _{z}.  \label{eq:con1}
\end{equation}

In this way, we arrive to an if and only if condition: If Eqs.(\ref{eq:con0},%
\ref{eq:con1}) is satisfied, then the tangential velocity of circular stable
equatorial orbits is constant. Furthermore, if the tangential velocity of
circular stable equatorial orbits is constant, then the metric coefficients
have to satisfy Eqs.(\ref{eq:con0},\ref{eq:con1}). Notice then that if the
function $\psi $ and $\mu $ are related by

\begin{equation}
e^{\psi }=({\frac{{\mu }}{{\mu _{0}}}})^{l}.  \label{eq:fin1}
\end{equation}

\noindent with $l=const,$ we obtain that this a necessary and sufficient
condition for the velocity ${{v_{c}}^{(\varphi )}}$ to be the same for two
orbits at different radii at the equatorial plane, provided that $l={({v_{c}}%
^{(\varphi )})^{2}/}\left( {1+({v_{c}}^{(\varphi )})^{2}}\right)$.\newline

We call your attention to the remarkable fact that the metric coefficient $%
\gamma $ does not play any role in this analysis, the motion analyzed is
determined only by the other two metric coefficients, which now are related
by this last equation, Eq.(\ref{eq:con1}), thus leaving the problems in
terms of only one metric coefficient. Actually this absence of $\gamma $
will be clear in the next section, where with the field equations we will
see that it is determined in terms of the other metric coefficients and some
components of the matter presented in the space-time, implying that it is
not and independent function.\newline

Thus, in order to have tangential velocities of equatorial objects circling
the galaxy, and whose magnitude is radii independent, the form of the line
element in the equatorial plane has to be

\begin{equation}
ds^{2}=-\left( {\frac{{\mu }}{{\mu _{0}}}}\right) ^{2l}\,dt^{2}+\left( {%
\frac{{\mu }}{{\mu _{0}}}}\right) ^{-2l}\,[e^{2\,\gamma }\,d\rho ^{2}+\mu
^{2}\,d{\varphi }^{2}].  \label{rcmetric}
\end{equation}

Notice that this type of space time definitely can not be asymptotically
flat. Neither it has the form of a space time related with a central black
hole. What can be said is that this line element describes the region where
the tangential velocity of the test particles is constant all over that
region, and that it has to be joined in the interior and in the exterior
regions with other types of space times if one wishes to have a central
black hole, and that the influence of the middle region ends at some
distance and thus has an asymptotically flat external region.\newline

Taking into account the constraint between $\psi$ and $\mu$ given by Eq.(\ref
{eq:fin1}), the energy, angular momentum, the rotational velocity , and the
second derivative of the potential have the final expressions: 
\begin{eqnarray}
E &=&{\frac{\left( {{\frac{{\mu }}{{\mu _{0}}}}}\right) {^{l}}}{\sqrt{l_{-}}}%
}, \\
L &=&{\frac{{\mu _{0}\,{v_{c}}^{(\varphi )}}\left( {\frac{{\mu }}{{\mu _{0}}}%
}\right) ^{-{1/l}_{+}}}{\sqrt{l_{-}}}}, \\
\Omega &=&{\frac{{{v_{c}}^{(\varphi )}}}{{\mu _{0}}}}\,\left( {\frac{{\mu }}{%
{\mu _{0}}}}\right) ^{-l_{-}/{l}_{+}}, \\
{\frac{{\partial ^{2}V(\rho )}}{{\partial \,\rho ^{2}}}}|_{extr}
&=&2\,e^{2(\psi -\gamma )}\,{\frac{{l}_{+}{-l}_{-}}{{l}_{+}^{2}}}\left( {%
\frac{{\mu _{\rho }}}{{\mu }}}\right) ^{2}.
\end{eqnarray}
being $l_{+}=1+{({v_{c}}^{(\varphi )})^{2}}$ and $l_{-}=1-{({v_{c}}%
^{(\varphi )})^{2}}$.\newline

Notice that the second derivative of the potential at the extreme is always
positive, thus the circular equatorial curves with constant tangential
velocity are stable.\newline

Before going to the field equations, we think it useful to present our
derivations applied to the Schwarzschild case, thus testing the expressions
while recovering the well known results. Starting with the line elements in
spherical coordinates: 
\begin{equation}
ds^{2}=-\left( 1-{\frac{{2M}}{{r}}}\right) \,dt^{2}+\left( 1-{\frac{{2M}}{{r}%
}}\right) ^{-1}\,dr^{2}+r^{2}\,(d\theta ^{2}+\sin ^{2}\theta \,d\phi ^{2}),
\end{equation}

\noindent we perform the coordinate transformation

\begin{equation}
r=\sqrt{\rho ^{2}+z^{2}}+{\frac{{M^{2}}}{{4\,\sqrt{\rho ^{2}+z^{2}}}}}%
+M,\,\,\,\theta =\tan ^{-1}({\frac{{\rho }}{{z}}}),
\end{equation}

\noindent (the inverse transformation is $\rho =R\,\sin \theta ,z=R\,\cos
\theta $, with $R={\frac{{1}}{{2}}}\,(r-M+\sqrt{r^{2}-2\,M\,r})$), to obtain
the line element in the Papapetrou form, Eq.(\ref{eq:ele}), with $e^{2\psi
}=\left( {\frac{{\sqrt{\rho ^{2}+z^{2}}-{\frac{{M}}{{2}}}}}{{\sqrt{\rho
^{2}+z^{2}}+{\frac{{M}}{{2}}}}}}\right) ^{2},e^{2\gamma }=\left( 1-{\frac{{%
M^{2}}}{{4\,(\rho ^{2}+z^{2})}}}\right) ^{2},\mu =\rho \,\left( 1-{\frac{{%
M^{2}}}{{4\,(\rho ^{2}+z^{2})}}}\right) ,\omega =0$. The horizon in this
coordinates is located at $\sqrt{\rho ^{2}+z^{2}}={\frac{{M}}{{2}}}$.\newline

Restricting the expressions to the equatorial plane, $z=0$, we have that $%
\psi |_{z=0}=\ln \left( \left( {\rho -{\frac{{M}}{{2}}}}\right) /\left( {%
\rho +{\frac{{M}}{{2}}}}\right) \right) ,$ $\mu |_{z=0}=\rho \,\left( 1-{%
M^{2}/4\,\rho ^{2}}\right) $, thus from Eqs.(\ref{eq:EL1}), we obtain, for
the energy and angular momentum:

\begin{equation}
E={\frac{\left( {\rho -{\frac{{M}}{{2}}}}\right) {^{2}}}{\left( {\rho +{%
\frac{{M}}{{2}}}}\right) {\,\sqrt{\rho ^{2}-2M\rho +{\frac{{M^{2}}}{{4}}}}}}}%
,\,\,\,L=\left( \rho +{\frac{{M}}{{2}}}\right) ^{2}\,\sqrt{{\frac{{M}}{{\rho
\,(\rho ^{2}-2M\rho +{\frac{{M^{2}}}{{4}}})}}}},
\end{equation}

\noindent for the angular velocity test particles, from Eq.(\ref{eq:om2}):

\begin{equation}
\Omega =\sqrt{M}\,\rho ^{-{\frac{{3}}{{2}}}}\,\left( 1+{\frac{{M}}{{2\rho }}}%
\right) ^{-3},
\end{equation}

\noindent which gives us the Kepler law.. For the tangential velocity, from
Eq.(\ref{eq:vphig}) we obtain

\begin{equation}
v^{(\varphi )}=\sqrt{{\frac{{M}}{{\rho }}}}\,\left( 1-{\frac{{M}}{{2\rho }}}%
\right) ^{-1},
\end{equation}

\noindent with the known dependence as the inverse of the square root of the
distance. Transforming back to spherical coordinates, it can be seen that
our expressions agree with the usual ones, see for instance \cite{Lig}. For
the second derivative of the potential, it is useful to write $\rho $ in
terms of the horizon radius, as $\rho =n\,{\frac{{M}}{{2}}}$, with $n$ a
number, and from Eq.(\ref{eq:potg}) we obtain:

\begin{equation}
V_{\rho \rho }|_{extr}={\frac{{4\,e^{2(\psi -\gamma )}}}{{M^{2}\,n\,(n+1)^{3}%
}}}\,{\frac{{(n^{2}-10n+1)}}{{(n^{2}-4n+1)}}}.
\end{equation}

This second derivative is positive down to $\rho =(5+2\sqrt{6}){\frac{{M}}{{2%
}}}$, which marks the last stable orbit, and corresponds to the known result
of $r=6M$ in spherical coordinates.\ 

In this way, we are confident on our expressions and can proceed to study
the field equations.\newline

A last remark about the geometric analysis. Recall that the observations are
based on measurements of the red shift, not on the tangential velocity
directly. If the space time is flat, the two quantities are proportional.
But we are now working in curved space times, so we have to see if that
proportionality is still valid. Following \cite{DSU}, we use the fact that
the frequency of a photon is given by $\nu =u^{\alpha }\,p_{\alpha }$, with $%
u^{\alpha }$ the four velocity of the object and $p^{\alpha }$ the photon
momentum, we have that the red shift, $z$, is given by

\begin{equation}
z=1-{\frac{{\nu _{em}}}{{\nu _{rec}}}},
\end{equation}

\noindent thus, for an object orbiting the galactic center in the equatorial
plane at a distance $\rho $ from the center, with tangential velocity $%
v^{(\varphi )}$ and emitting a photon with frequency $\nu _{0}$, and for an
observer located at rest at infinity, that is far away from the emission,
detecting the photon with a frequency $\nu _{\infty }$, it can be shown that
the red shift is given by:

\begin{equation}
z=1-{\frac{{(1+v^{(\varphi )})}}{\sqrt{1-{v^{(\varphi )}}^{2}}}}\,\sqrt{{%
\frac{{g_{tt}(\rho )}}{{g_{tt}(\infty )}}}}.
\end{equation}

Now, for the observed velocities, we have that $v\ll 1$, {\it i. e.} they
are much less than the speed of light, and we have to suppose that far away
from the observed galaxy its gravitational influence ends, otherwise we
could not detect the tangential velocity, we would then be moving along with
the observed object! Thus we can take $g_{tt}(\infty )=-1$ and we have that
at first order in the velocity, with $g_{tt}(\rho )=-e^{\psi }=-1-\psi
-\cdots $, we obtain:

\begin{equation}
z=-(v^{(\varphi )}+\psi +\cdots ).
\end{equation}

But we have computed that in the case analyzed $\psi ={{v^{(\varphi )}}^{2}/}%
\left( {1+{v^{(\varphi )}}^{2}}\right) \,\ln ({\mu /\mu _{0}})\sim 2\,{%
v^{(\varphi )}}^{2}\,({\mu -1)/(\mu +1})$, thus $\psi \sim {v^{(\varphi )}}%
^{2}$, and we conclude that $z\sim v^{(\varphi )}$. In this way, we see that
the radii independence of the value of the measured red shift can be related
with the radii independence of the value of the tangential velocity, which
is the fact that has been studied in this work.\newline

\section{Field equations}

Now we are in a position where we can test any type of matter-energy to
determine whether or not it produces a curvature in the space time such that
the motion of the test particles can be circular stable and be such that the
tangential velocity of those particles is constant for a large radial region
in the equatorial plane.\newline

We obtain the general form of the Einstein tensor for the axisymmetric
static space time described by Eq. (\ref{eq:ele}), with $\omega =0$ , and
equate it to an arbitrary stress energy tensor. After some manipulations we
conclude that the field equations are a set of two equations involving the
metric coefficients $\psi $ , and $\mu $:

\begin{eqnarray}
&&\mu(\psi_{\rho\rho}+\psi_{zz}) + \mu_\rho\,\psi_\rho+ \mu_z\,\psi_z=4\pi
\,\mu \,[e^{-2(\psi -\gamma )}\,(e^{-2\psi }T_{tt}+{\frac{{e^{2\psi }}}{{\mu
^{2}}}}\,T_{{\varphi }{\varphi }})+T_{\rho \rho }+T_{zz}],  \label{ee1} \\
&&\mu_{\rho\rho}+\mu_{zz} =8\pi \,\mu \,[T_{\rho \rho }+T_{zz}].  \label{ee2}
\end{eqnarray}

\noindent There are also two first order equations for the other metric
coefficient $\gamma $:

\begin{eqnarray}
&&\gamma _{\rho }\,\mu _{\rho }-\gamma _{z}\,\mu _{z}-\mu \,({\psi _{\rho }}%
^{2}-{\psi _{z}}^{2})+\mu _{zz}=8\,\pi \,\mu \,T_{\rho \rho }, \\
&&\gamma _{\rho }\,\mu _{z}+\gamma _{z}\,\mu _{\rho }-2\,\mu \,\psi _{\rho
}\,\psi _{z}-\mu _{\rho z}=8\,\pi \,\mu \,T_{\rho z},
\end{eqnarray}

\noindent and finally, the field equations give us another equation for the
second derivatives of $\gamma $ which thus is redundant, this equation is:

\begin{equation}
\gamma_{\rho\rho}+\gamma_{zz} +(\psi_\rho)^{2}+ (\psi_z)^{2}=8\,\pi \, {%
\frac{{e^{2\gamma }}}{{\mu ^{2}}}}\,T_{{\varphi }{\varphi }}.  \label{eee1}
\end{equation}

The analysis presented in the last section is exact and the relations
between the metric coefficients and their first derivatives must be
satisfied at the equatorial plane in order to describe the observed motion.
Using the Einstein's equations, we need the second derivatives of those
metric coefficients. Thus, we have to make the approximation that the
relations obtained among them, holds as well in a region close to the
equatorial plane. Within this approximation, from Eqs.(\ref{eq:con0}, \ref
{eq:con1},\ref{eq:fin1}), it can be obtained the following expression: 
\begin{equation}
\mu(\psi_{\rho\rho}+\psi_{zz}) + \mu_\rho\,\psi_\rho+ \mu_z\,\psi_z ={({v_{c}%
}^{(\varphi )})^{2}/}\left( {1+({v_{c}}^{(\varphi )})^{2}}\right)
\,(\mu_{\rho\rho}+\mu_{zz}).  \label{eq:2der}
\end{equation}
Thus, with this last relation, from the Einstein's equations Eqs. (\ref{ee1},%
\ref{ee2}) we obtain a constraint among the stress energy tensor components
which, within the approximation made on the validity of Eqs.(\ref{eq:con0}, 
\ref{eq:con1},\ref{eq:fin1}) out of the equatorial plane, has to be
satisfied by any type of matter in order to have constant tangential
velocities: 
\begin{equation}
-\left( {\frac{{1-({v_{c}}^{(\varphi )})^{2}}}{{1+({v_{c}}^{(\varphi )})^{2}}%
}}\right) (T_{\rho \rho }+T_{zz})=e^{-2(\psi -\gamma )}\left( e^{-2\psi
}T_{tt}+{\frac{{e^{2\psi }}}{{\mu ^{2}}}}\,T_{{\varphi }{\varphi }}\right) ,
\label{eq:fin}
\end{equation}

We finish our analysis by testing several types of matter, described by
their respective stress energy tensor, to see whether or not they are able
to deform the geometry of the space time in such a way that the tangential
velocity of the equatorial rotational objects be constant, that is, that
they satisfy Eq.(\ref{eq:fin}).\newline

\subsection{Vacuum Fields}

We start with the vacuum solutions, with $T_{\mu \nu }=0$. In this case Eq.(%
\ref{eq:fin}) is trivially satisfied, thus we proceed to analyze the
Einstein equations directly. From Eq.(\ref{ee2}), the easiest solution
implies $\mu =\rho $. Eq.(\ref{ee1}), is a Laplace equation for $\psi $ and
`imposing the condition that at the equatorial plane the flat curve
condition, Eq.(\ref{eq:con1}), be satisfied, as well as the one of symmetry
with respect to the galactic plane, we obtain that $\psi =l\,\ln {\rho }$
and the other Einstein equation then imply $\gamma =l^{2}\,\ln {\rho }$,
with $l={({v_{c}}^{(\varphi )})^{2}/}\left( {1+({v_{c}}^{(\varphi )})^{2}}%
\right) $. In this way, we obtained an exact vacuum solution for the
Einstein equations, which produces that the test particles circling at the
equatorial plane behave in agreement with the observations: 
\begin{equation}
ds^{2}=-\rho ^{2l}\,dt^{2}+\rho ^{-2l}(\rho ^{2l^{2}}\,(d\rho
^{2}+dz^{2})+\rho ^{2}\,d{\varphi }^{2}).  \label{civita}
\end{equation}

The central object is string-like. Observations show that cosmic strings
object are very unlikely to exist, nevertheless, this is an example of
objects which could produce the observed motion of test particles, and in
which the density does not go as $r^{-2}$, because it is vacuum. Thus, such
a behavior on the density is a sufficient but not a necessary condition for
the flatness of the rotational curves.\newline

\subsection{Perfect Fluid}

\label{subsec:pf}

For the perfect fluid, $T_{\mu \nu }=(d+p)\,u_{\mu }\,u_{\nu }+g_{\mu \nu
}\,p$, with $d$ the density of the fluid and $p$ its pressure. In this case
we are thinking on a ``dark fluid'', which could be composed of planetoids
or WIMPS or MACHOS, which are not seen but it is thought that they could be
there affecting the geometry in the way needed in order to have the observed
behavior in the tangential velocities of the luminous matter. Taking this
dark fluid as static, the four velocity is given by $u^{\alpha
}=(u^{0},0,0,0)$ with, for the line element given by Eq.(\ref{eq:ele}) with $%
\omega =0$, $u^{0}=E\,e^{-2\psi }$, and $L=0$. Thus $u_{0}=-E$, and from $%
u_\alpha\,u^\alpha=-1$, we conclude that $E=e^{\psi}$. The stress energy
tensor for the dark fluid then has the form: 
\begin{eqnarray}
&&T_{tt}=d\,E^{2}=d\,e^{2\,\psi} \\
&&T_{\rho \rho }=T_{zz}=e^{-2(\psi -\gamma )}p, \\
&&T_{{\varphi }{\varphi }}=\mu^{2}\,e^{-2\psi }p.
\end{eqnarray}
Substituting in Eq.(\ref{eq:fin}), we obtain that in the equatorial plane,
in order to satisfy the observed behavior on the tangential velocities, the
``dark fluid'' has to satisfy:

\begin{equation}
-2\left( {\frac{{1-({v_{c}}^{(\varphi )})^{2}}}{{1+({v_{c}}^{(\varphi )})^{2}%
}}}\right) \,p=(d+p),
\end{equation}

Thus, we obtain an equation of state for the ``dark fluid'' particles at the
equatorial plane: 
\begin{equation}
p=-{\frac{{1+({v_{c}}^{(\varphi )})^{2}}}{{3-({v_{c}}^{(\varphi )})^{2}}}}%
\,d,  \label{eqstate}
\end{equation}
which, compared to the equation of state for a perfect fluid, $p=\omega \,d$%
, implies that $-1<\omega <-{\frac{1}{{3}}}$, for ${v_{c}}^{(\varphi )}$
between the speed of light and zero. This result is quite remarkable. It
coincides with the type of equation of state derived within the Quintessence
model \cite{Quin0,Quin,Quin2,Quin3} at the cosmological level, and now we
obtain similar results at the galactic level. This sort of matter has been
called exotic matter \cite{Thor} and studied in several contexts \cite
{goodman}. Our result points to the fact that the Dark Matter actually could
be exotic. We want to stress that due to the approximation taken for the
behavior of the metric coefficients off the galactic plane, we are not
excluding the possibility that the dark fluid be composed of baryonic usual
matter, actually from the Newtonian approach, we know that regular matter
can produce the observed motion. This fact is not reproduced in the present
analysis, due to our approximation, what we certainly can conclude is that
exotic type of matter also can produce such `observed motion.\newline

In order to recover the Newtonian case, where we know that the dust type
fluid does work as the dark matter, we have to analyze the spherical case,
which is introduced in Appendix \ref{app:1}. For the stress energy tensor we
again take the static perfect fluid, thus the four velocity of the test
particle reads $u^\mu=(u^0,0,0,0)$, with $u^0=\dot{t}$. For the spherically
symmetric metric, Eq.(\ref{metric_sp}), we know that $\dot{t}$ is associated
to a conserved quantity, the energy, and is given by: $\dot{t}={\frac{E}{{%
B(r)}}}$, and from the normalization of the four velocity, $%
u_\alpha\,u^\alpha=-1$, we get that $E=\sqrt{B(r)}$. Thus, considering those
spacetimes for which the tangential velocity ($v_{c}$) of test particles in
circular orbits is radii independent, we use Eq.(\ref{eq:Bes}), obtaining
that $u^0={\frac{1}{\sqrt{B_0}}}\,r^{-{(v_{c})}^2}$, thus, $u_0=-\sqrt{B_0}%
r^{{(v_{c})}^2}$. In this way, we get that the non zero components of the
static spherically symmetric perfect fluid are:

\begin{eqnarray}
T_{tt}&=&d\,B_0\,r^{2\,{(v_{c})}^2},  \nonumber \\
T_{rr}&=&p\,A(r),  \nonumber \\
T_{\theta\theta}&=&p\,r^2.
\end{eqnarray}

Substituting these expressions in the Einstein equations Eq.(\ref{eq:EEsp}),
it turns out that the equations can be completely solved, yielding: 
\begin{eqnarray}
A(r)&=&{\frac{{b}}{{2(1-A_0\,r^{\frac{{b}}{{a}}})}}}, \\
d&=&{\frac{{(2+a)\,A(r)-a-b}}{{8\,\pi\,a\,r^2\,A(r)}}}, \\
p&=&{\frac{{1+2\,{(v_{c})}^2-A(r)}}{{8\,\pi\,r^2\,A(r)}}},
\end{eqnarray}
with $b=2\,(1+2\,{(v_{c})}^2-{v_{c}}^4), a=1+{(v_{c})}^2$, and $A_0$ an
integration constant.\newline

Taking the particular case for the integration constant $A_0=0$, we get

\begin{eqnarray}
A(r)&=&{\frac{{b}}{{2}}}, \\
d&=&{\frac{{{(v_{c})}^2\,(1-{\frac{{{(v_{c})}^2}}{{2}}})}}{{2\,\pi\,b\,r^2}}}%
, \\
p&=&{\frac{{{(v_{c})}^4}}{{4\,\pi\,b\,r^2}}}.
\end{eqnarray}

In this way, we obtain the particular solution where the particles move in
the observed way, in a space-time with a deficit angle, $g_{rr}=constant$,
and in which the density goes as $d\sim {{(v_{c})}^{2}/r^{2}}$, and the
pressure goes as $p\sim {{(v_{c})}^{4}/r^{2}}$, thus it is a dust like
solution. Furthermore, for an equation of state $p=\omega d$, this dust like
solution implies $\omega ={{(v_{c})}^{2}/}\left( {2\,(1-{(v_{c})}^{2}/2)}%
\right) $, which is between $0$ and $1$, for the tangential velocity between 
$0$ and $1$, and thus is a perfectly well known fluid type. In this way we
recover the dust hypotheses within our approach, and we clearly see that
the` fact that we did not recover this case within the axi-symmetric
analysis, was due to our approximation outside the equatorial plane.\newline

For the general case when the integration constant $A_{0}$ is non zero,
taking an equation of state as before $p=\omega d$, we obtain that the $%
\omega $ is a function of $r$ and again, as in the axial case, it is
negative, so we are dealing with the exotic type of perfect fluid obtained
in the axi-symmetric case.\newline

This is a good moment \ to discuss the question on the need to use GR, even
though the gravitational field is weak. In the Newtonian description it is
well known that the space-time can be described as 

\[
ds^{2}=-(1+2\Phi )dt^{2}+(1-2\Phi )dr^{2}+r^{2}d\Omega ^{2}
\]
with $\Phi =\Phi (r)$ the Newtonian gravitational potential. For the
spherically symmetric case we obtain that 

\[
-g_{tt}=r^{2\left( v_{c}\right) ^{2}}=e^{2\left( v_{c}\right) ^{2}\ln
(r)}=1+2\left( v_{c}\right) ^{2}\ln (r)+\cdots 
\]
from here we determine that $\Phi =\left( v_{c}\right) ^{2}\ln (r)$. In this
Newtonian approximation the complete set of 10 Einstein equations reduces to
one equation, the usual Poisson equation $\nabla ^{2}\Phi =4\pi Gd$.
Collecting this last result, it is obtain that the density, $d,$ goes as $%
d\sim \left( v_{c}\right) ^{2}/r^{2}$, which is the expression for the dark
matter density known from the astronomers' work.

However, notice that in this approximation it can not be said anything else
about the matter producing the observed motion. The Newtonian approximation 
{\bf fixes} the matter to be dust-perfect-fluid-like type. This is the usual
way of reasoning: it is suppose a priori that the dark matter is a
completely Newtonian dust and at the end of the day one arrives to a
consistent description of the dark matter determining only the shape of the
Newtonian gravitational potential. 

In this work we are proceeding in a different way. We are using Einstein
equations backwards; we do not make any assumptions in the type of matter
nor do approximations. From the observations on the motion of the test
particles, we determine the geometry and then, by means of the Einstein's
equations we obtain constrains of the type of matter. We have shown that we
do recover the Newtonian result, but also it is clear that this is a very
particular case for a very specific type of matter. In the general
reasoning, we do not fix neither the type of matter nor the equation of
state, we let the equations themselves to do that obtaining more general
results.

To end this argumentation we recall the reader that there is {\bf three
conditions} that have to be fulfill in order to reach the Newtonian limit;
1) the speeds of the study particles most be most less than the speed of
light, 2) the gravitational field must be weak, {\bf and }3) the pressures
associated with the matter study must to be must smaller than the
corresponding density. It is this last condition which is taking a priori in
the usual analysis and it is not satisfy in the general case as we have
shown, thus justify the need of GR in order to be able to consider any type
of matter.

As a last remark about these results on the type of perfect fluid is that
the Big-Bang nucleosynthesis imposes very strong constraints to the
percentage of the baryonic matter to the total content of the Universe. If
the dark matter would be a dark fluid of baryonic matter, such percentage
would be quite above the value settled by those constraints. Thus, even if
the baryonic dark fluid can not be discarded by dynamical methods, the
cosmological constraints make it unlikely, a fact which might strength the
case for exotic type of dark fluids or for other type of dark matter.\newline

\subsection{Cosmological Constant}

For a cosmological constant, $\Lambda $, $T_{\mu \nu }=\Lambda \,g_{\mu \nu }
$. We thus have $T_{tt}=-e^{2\psi }\Lambda ,T_{\rho \rho}=T_{zz}=e^{-2(\psi
-\gamma )}\,\Lambda ,T_{{\varphi }{\varphi }}=e^{-2\psi }\,\mu ^{2}\,\Lambda 
$, thus from Eq.(\ref{eq:fin}), we obtain

\begin{equation}
\left( {\frac{{1-({v_{c}}^{(\varphi )})^{2}}}{{1+({v_{c}}^{(\varphi )})^{2}}}%
}\right) \,\Lambda =0,
\end{equation}

In this way we see that, within our approximation, a non zero cosmological
constant can not explain the observed behavior because implies that the
observed tangential velocity had to be equal to 1, {\it i. e.}, they should
be moving at the speed of light. Something similar occurs with the scalar
field.\newline

\subsection{Scalar Field}

For scalar field $\phi $ with potential, $T_{\mu \nu }=\phi _{,\mu }\phi
_{,\nu }-{\frac{{1}}{{2}}}\,g_{\mu \nu }\,\phi ^{\alpha }\,\phi _{\alpha
}+g_{\mu \nu }\,V(\phi )$. We have that, due to the symmetry of our space
time $\phi =\phi (\rho ,z)$ the $T_{\mu \nu }$ components are:

\begin{eqnarray}
T_{tt} &=&{\frac{{1}}{{2}}}e^{2(2\psi -\gamma )}\,({\phi _{\rho }}^{2}+{\phi
_{z}}^{2})-e^{2\psi }\,V(\phi ), \\
T_{\rho \rho } &=&{\frac{{1}}{{2}}}\,({\phi _{\rho }}^{2}-{\phi _{z}}%
^{2})+e^{-2(\psi -\gamma )}\,V(\phi ), \\
T_{zz} &=&-{\frac{{1}}{{2}}}\,({\phi _{\rho }}^{2}-{\phi _{z}}%
^{2})+e^{-2(\psi -\gamma )}\,V(\phi ), \\
T_{\rho z} &=&\phi _{\rho }\,\phi _{z}, \\
T_{{\varphi }{\varphi }} &=&-{\frac{{1}}{{2}}}\,e^{-2\gamma }\,\mu ^{2}({%
\phi _{\rho }}^{2}+{\phi _{z}}^{2})+e^{-2\psi }\,\mu ^{2}\,V(\phi ).
\end{eqnarray}

Inserting these components in Eq.(\ref{eq:fin}), we obtain that, as in the
cosmological constant case

\begin{equation}
\left( {\frac{{1-({v_{c}}^{(\varphi )})^{2}}}{{1+({v_{c}}^{(\varphi )})^{2}}}%
}\right) \,V(\phi )=0.
\end{equation}

Again, either the particles move at the speed of light, or the scalar field
potential is zero at the equatorial plane. When the scalar field potential
is zero, that is, we have a massless scalar field, Eq.(\ref{eq:fin}) is
satisfied, and we have to go back to the Einstein's equations as in the
vacuum case. Again from Eq.(\ref{ee2}), we take the simplest solution $\mu
=\rho $, and, as in the vacuum case, we obtain $\psi =l\,\ln {\rho }$. The
last metric coefficient $\gamma $, can be solved in terms of the scalar
field. Thus, to have a complete solution, it is only left to solve the Klein
Gordon equation for the scalar field: $D^{2}\phi +{\frac{{1}}{{\mu }}}D\phi
\,D\mu =0$, which turns out to be the same equation for the metric
coefficient $\psi $. However, in this case we do not have boundary
conditions well defined: The space time is not asymptotically flat; it is
not known the form of the space time near and at the origin, we are only
analyzing the region where the curves are flat; there are no conditions for
the scalar field at the equatorial plane. What can be concluded at this
stage, is that the scalar field does remain being a candidate for the dark
matter, and thus to contribute with about 25\% of the matter of the Universe.%
\newline

\section{Conclusions}

We have found that in a static, axisymmetric space-time, a sufficient and
necessary condition in order to have a flat profile for the rotational
curves in a plane of that space-time is that its metric tensor must have the
form given by Eq. (\ref{rcmetric}). This form of the metric must be the one
required for galaxies in the region where the rotational curves profile of
stars is flat. It is important to stress the fact that in the derivation of
this expression, only the geometry of the space-time was involved, thus, it
is independent of the type of matter which generates such a geometry; that
is, whatever the matter might be, Eq. (\ref{rcmetric}) must be the form of
the line element at the galactic plane for a static axisymmetric space time
which can be expressed in the Papapetrou form, Eq. (\ref{eq:ele}) with $%
\omega =0$, thus, this result is not only the general relativistic analog of
the Newtonian result for the gravitational Newtonian potential, $\phi
(r)\sim 1/r$, but it can be used for any type of matter, including those
which do not have a clear Newtonian expression, such as the scalar field.%
\newline

With this idea in mind, we proceeded further using the Einstein's equations,
which essentially describe the inter dependence of matter-energy and
geometry. We had to accept some loose of the generality of our results, in
making the assumption that the definite relation which we obtained for the
metric coefficients at the galactic plane, Eqs.(\ref{eq:fin1}), are also
valid in a close by region off the plane, and thus determine a relation for
the second derivatives of the metric coefficients, Eq.(\ref{eq:2der}).
Within this approximation, we were able to obtain a constraint equation
among the components of a general stress energy tensor, $T^{\alpha \beta },$
Eq.(\ref{eq:fin}). We tested this expression in four types of stress energy
tensors which included the traditional types of matter which have been used
as candidates for the dark matter in the galactic halos, such as the perfect
fluid or the cosmological constant. We obtained that for the vacuum case, a
cosmic string type of matter does generates the observed motion of test
particles. Even though the cosmic strings are unlikely objects to be the
Universe, it was a clear example for the fact that the Newtonian behavior of
the density, $\rho (r)\sim 1/r^{2}$, is not a necessary condition for
describing the observed motion. We analyzed also the static perfect fluid,
and it is interesting that in difference with the Newtonian description,
within the general relativistic formulation, we are able to obtain
conditions for the equation of state of the dark fluid. Even though we were
not able to reproduce the well known Newtonian result for dust-like fluid,
due to our approximation for the second derivatives of the metric
coefficients, we did showed that a dark fluid with an exotic type of matter,
is a candidate for being the dark matter. Furthermore, this results
represents the Quintessence type of matter at a galactic level. We performed
the study in the spherically symmetric static space time in appendix \ref
{app:1}, and when applied to the perfect fluid static case we did recover
the Newtonian result, namely that the ``dark fluid'' could be a well behaved
dust like fluid. We further analyzed the cosmological constant case, which
within our approximation implied that it can not be, and the massive scalar
field case, which again, within our approximation, turned out that it has to
be massless, and the massless scalar field also remains being a candidate
for the dark matter.\newline

As we have mentioned above, our results are useful for describing the region
where is observed the flat behavior of the test particles rotating around
the galactic center, it is clearly needed to proceed further in order to be
able to describe the motion in the complete region, from the center to the
exterior. Some preliminary results indicate that a combination of perfect
fluid with baryonic matter and some of the matter analyzed here, could be in
the right direction, \cite{tofc}. Also, the approximation we made for the
second derivatives of the metric coefficients, has to be further analyzed.
It would be of great help, in order to be able to apply the present
description in objets out from the glactic plane, to have a sample of the
profile of velocities of such objects. In the simpler spherically symmetric
case, there is no need to make such approximation, and more definite results
can be obtained, though with more restrictions on the geometry. Besides,
there are reasons to belive that the dark matter halo is spherical \cite
{Ibata}, thus it might be a good approximation the analysis made within this
symmetry. We studied the scalar field in this symmetry in \cite{scalar}.%
\newline

Finally, as the best way to study dark matter is through its effects on the
dynamics of the visible objetcs, further studies along the lines presetned
in this work can be performed using gravitational lensing or jets.

Overall, we consider that the analysis presented in this work, is on the
right track in order to determine which is the type of matter which
constitutes the 90\% of the matter in the galaxies.\newline

\section{Acknowledgments}

This work was partially supported by a grant CONACyT-DFG, by CONACyT,
M\'{e}xico, under grants 94890 (F.S.G.), and DGAPA-UNAM IN121298 (D.N.) We
want to thank the relativity group in Jena for its kind hospitality and
partial support.\newline

\appendix

\label{app:1}

In this appendix we present an analogous derivation for the conditions which
the constancy of the tangential velocity of circular orbits impose on the
metric coefficients for the spherical static case. Furthermore, we also
present the Einstein equations in this case for a general stress energy
tensor.\newline

It is interesting that, due to the symmetries, in this case we do not have
to restrict the analysis to equatorial orbits, and that the above mentioned
conditions give a closed form for the metric coefficient $g_{tt}$.\newline

We begin with the line element 
\begin{equation}
ds^{2}=-B(r)dt^{2}+A(r)dr^{2}+r^{2}(d\theta ^{2}+\sin^{2}\theta d\varphi^{2})
\label{metric_sp}
\end{equation}

The Lagrangian for a test particle reads 
\begin{equation}
2{\cal {L}}=-B(r)\dot{t}^{2}+A(r)\dot{r}^{2} +r^2\,(\dot{\theta}^2 +
\sin^{2}\theta\,\dot{\varphi}^{2}).  \label{eq:lages}
\end{equation}
We infer the conserved quantities, the energy $E=B(r)\dot{t}$, the $\varphi$%
-momentum $L_\varphi=r^2\,\sin^{2}\theta\dot{\varphi}$, and the total
angular momentum, $L^2={L_\theta}^2+({\frac{{L_\varphi}}{{\sin\theta}}})^2$,
with $L_\theta=r^2\dot{\theta}$. The radial motion equation can thus be
written as: 
\begin{equation}
\dot{r}^{2}+V(r)=0,  \label{eq:rades}
\end{equation}
with the potential $V(r)$ given by 
\begin{equation}
V(r)=-{\frac{{1}}{{A(r)}}}({\frac{{E^2}}{{B(r)}}}-{\frac{{L^2}}{{r^2}}}-1).
\label{eq:potes}
\end{equation}

Notice that, due to the spherical symmetry, we do not need to restrict the
study to equatorial orbits, this last radial motion is valid for any angle $%
\theta$. For circular stable orbits, we again have the conditions, $\dot{r}%
=0, V_r=0$, and $V_{rr}>0$, which imply the following expressions for the
energy and total momentum of the particles in such orbits:

\begin{eqnarray}
E^{2} &=&{\frac{{2B(r)^{2}}}{{2B(r)-rB(r)_{r}}}}, \\
L^{2} &=&{\frac{{r^{3}B(r)_{r}}}{{2B(r)-rB(r)_{r}}}},  \label{eq:ELes}
\end{eqnarray}

\noindent and for the second derivative of the potential evaluated at the
extrema

\begin{equation}
V(r)_{rr}|_{extr} =2{\frac{{{\frac{{rB(r)_{rr}}}{{B}}}+{\frac{{B(r)_{r}}}{{B}%
}} (3-{\frac{{2rB(r)_{r}}}{{B}}})} }{{rA(r)(2-{\frac{{rB(r)_{r}}}{{B}}} )}}}.
\end{equation}

On the other hand, in a similar manner as it was presented in the text, we
obtain that the tangential velocity, ${(v_{c})}^2={\frac{{r^2}}{{B(r)}}}(({%
\frac{{d\theta}}{{dt}}})^2 +\sin^2\theta({\frac{{d\varphi}}{{dt}}})^2)$, for
particles in stable circular orbits is given by:

\begin{equation}
{(v_{c})}^{2}={\frac{{rB(r)_{r}}}{{2B(r)}}}.
\end{equation}

Thus, imposing the observed condition that this tangential velocity velocity
is constant for several radii, this last equation can be integrated for the
metric coefficient $g_{tt}$:

\begin{equation}
B(r)=B_0\,r^{2\,{(v_{c})}^2},  \label{eq:Bes}
\end{equation}

\noindent with $B_0$ an integration constant.\newline

In this way, we again arrive to a theorem, stating that: For a static
spherically symmetric spacetime, $v_{c}$, the tangential velocity of
particles moving in circular stable orbits is radii independent if and only
if the $g_{tt}$ metric coefficient has the form $g_{tt}=B_0\,r^{2\,{(v_{c})}%
^2}$.\newline

Notice that in this case, one of the metric coefficients was completely
integrated and the other one, $A(r)$ remains arbitrary. Also, as mentioned,
the analysis made no suppositions on the plane of motion, so the result is
valid for any circular stable trajectory.\newline

Finally, we can construct the Einstein tensor and arrive to the following
Einstein equations which give us information about the type of matter
curving the spacetime in such a way that the motion corresponds to the
observed one:

\begin{eqnarray}
\frac{B_{0}r^{2{(v_{c})}^{2}}}{A(r)^{2}}(rA(r)^{\prime }+A(r)(A(r)-1))
&=&8\,\pi \,T_{tt},  \nonumber \\
\frac{2\,{(v_{c})}^{2}+1-A(r)}{r^{2}} &=&8\,\pi \,T_{rr},  \nonumber \\
-\frac{r\,({(v_{c})}^{2}+1)\,A(r)^{\prime }-2\,{v_{c}}^{4}\,A(r)}{2\,A(r)^{2}%
} &=&8\,\pi \,T_{\theta \theta },  \label{eq:EEsp}
\end{eqnarray}

\noindent where ${}^{\prime}$ stands for derivative with respect to $r$.
This study was applied for the scalar field in \cite{scalar}.\newline

\label{app:2}

In this appendix we present the generalization of the derivation of the
constraint equation among the metric coefficients, Eq.(\ref{eq:con1}), for
the stationary case, where $\omega \neq 0$, described by the line element 
\ref{eq:ele}.\newline

>From Eq.(\ref{eq:mta}), we express $\dot{t}$, and $\dot{\varphi}$. in terms
of $E,L$, and the metric coefficients as 
\begin{eqnarray}
\dot{t} &=&{\frac{{e^{2\psi }}}{{\mu ^{2}}}}\,[(\mu ^{2}\,e^{-4\psi }-\omega
^{2})\,E-\omega \,L], \\
\dot{\varphi} &=&{\frac{{e^{2\psi }}}{{\mu ^{2}}}}\,(\omega \,E+L).
\label{eq:pfi}
\end{eqnarray}
Using these equation in the constraints ones, Eqs.(\ref{eq:cons0}) we arrive
at: 
\begin{eqnarray}
\mu ^{2}\,e^{-2\psi }\,(1-e^{-2\psi }\,E^{2})+(\omega \,E+L)^{2} &=&0, 
\nonumber \\
-(e^{-2\psi })_{\rho }\,E^{2}+\left( {\frac{{e^{2\psi }}}{{\mu ^{2}}}}%
\right) _{\rho }\,(\omega \,E+L)^{2}+{\frac{{2\,e^{2\psi }}}{{\mu ^{2}}}}%
\,(\omega \,E+L)\,\omega _{\rho }\,E &=&0,
\end{eqnarray}
Solving for $E$ and $L$, we obtain: 
\begin{eqnarray}
E &=&e^{\psi }\,\sqrt{{\frac{{\cal {A}}}{{{\cal {B}}}}}},  \nonumber \\
L &=&{\frac{{\mu \,e^{-\psi }}}{\sqrt{{\cal {B}}}}}\,(\sqrt{{\cal {A}}-{\cal 
{B}}}-{\frac{{\omega \,e^{2\psi }}}{{\mu }}}\,\sqrt{{\cal {A}}}),
\label{eq:ELfi}
\end{eqnarray}
where 
\begin{eqnarray}
{\cal {A}} &=&2\,e^{-4\psi }\,(\mu _{\rho }-\mu \,\psi _{\rho })(\mu _{\rho
}-2\,\mu \,\psi _{\rho })-(\omega _{\rho })^{2}+  \nonumber \\
&&  \nonumber \\
&&\pm \omega _{\rho }\,\sqrt{(\omega _{\rho })^{2}-4\,\mu \psi _{\rho
}\,e^{-4\psi }\,(\mu _{\rho }-\mu \,\psi _{\rho })}, \\
{\cal {B}} &=&2\,e^{-4\psi }\,(\mu _{\rho }-2\,\mu \,\psi _{\rho
})^{2}-2\,(\omega _{\rho })^{2}.  \label{Aeq:AB}
\end{eqnarray}

For the second derivative of the potential $V(\rho )$ evaluated at the
extreme, we obtain: 
\begin{eqnarray}
V_{\rho \rho }|_{extr} &=&-{\frac{{2\,e^{2(\psi -\gamma )}}}{{{\cal {B}}}}}%
[(2\,\psi _{\rho \rho }-{\frac{{\mu _{\rho \rho }}}{{\mu }}}+3\left( {\frac{{%
\mu _{\rho }}}{{\mu }}}\right) ^{2}-4{\frac{{\mu _{\rho }}}{{\mu }}}\,\psi
_{\rho }+{\frac{{e^{4\psi }}}{{\mu ^{2}}}}(\omega _{\rho })^{2})\,{\cal {A}}
\nonumber \\
&&-(\psi _{\rho \rho }-{\frac{{\mu _{\rho \rho }}}{{\mu }}}+2\,(\psi _{\rho
})^{2}+3\left( {\frac{{\mu _{\rho }}}{{\mu }}}\right) ^{2}-4{\frac{{\mu
_{\rho }}}{{\mu }}}\,\psi _{\rho })\,{\cal {B}}+  \nonumber \\
&&(4\,\left( \psi _{\rho }-{\frac{{\mu _{\rho }}}{{\mu }}}\right) \,\omega
_{\rho }+\omega _{\rho \rho })\,{\frac{\sqrt{{\cal {A}}({\cal {A}}-{\cal {B}}%
)}}{{\mu }}}].
\end{eqnarray}

On the other hand, using Eqs.(\ref{eq:pfi}), and (\ref{eq:ELfi}), in the
expression for the angular velocity, Eq.(\ref{eq:om1}) we obtain that: 
\begin{equation}
\Omega ={\frac{{e^{2\psi }}}{{\mu }}\frac{\sqrt{{\cal {A}}-{\cal {B}}}}{{%
\sqrt{{\cal {A}}}-{\frac{{\omega \,e^{2\psi }}}{{\mu ^{2}}}}\,\sqrt{{\cal {A}%
}-{\cal {B}}}}}},  \label{Aeq:om2}
\end{equation}
where ${\cal {A}}$ and ${\cal {B}}$ are given by Eqs.(\ref{Aeq:AB}).\newline

As in the static case, Following Chandrasekhar \cite{Chan}, we rewrite the
line element given in Eq.(\ref{eq:ele}) as: 
\begin{eqnarray}
ds^{2} &=&-{\frac{{\mu ^{2}\,e^{-2\,\psi }}}{{\mu ^{2}\,e^{-4\,\psi }-\omega
^{2}}}}\,dt^{2}+e^{2\psi }\,(\mu ^{2}\,e^{-4\,\psi }-\omega ^{2})\,\left(
d\varphi -{\frac{{\omega }}{{\mu ^{2}\,e^{-4\,\psi }-\omega ^{2}}}}dt\right)
^{2}+  \nonumber \\
&&e^{-2(\psi -\gamma )}(d\rho ^{2}+dz^{2}),  \label{Aeq:ele1}
\end{eqnarray}
thus, in terms of the proper time, $d\tau ^{2}=-ds^{2}$, we have that 
\begin{eqnarray}
d\tau ^{2} &=&{\frac{{\mu ^{2}\,e^{-2\,\psi }}}{{\mu ^{2}\,e^{-4\,\psi
}-\omega ^{2}}}}\,dt^{2}\,[1-{\frac{{e^{4\,\psi }\,(\mu ^{2}\,e^{-4\,\psi
}-\omega ^{2})^{2}}}{{\mu ^{2}}}}\left( {\frac{{d\varphi }}{{dt}}}-{\frac{{%
\omega }}{{\mu ^{2}\,e^{-4\,\psi }-\omega ^{2}}}}\right) ^{2}+  \nonumber \\
&&-{\frac{{e^{2\,\gamma }\,(\mu ^{2}\,e^{-4\,\psi }-\omega ^{2})}}{{\mu ^{2}}%
}}\left( \left( {\frac{{d\rho }}{{dt}}}\right) ^{2}+\left( {\frac{{dz}}{{dt}}%
}\right) ^{2}\right) ],
\end{eqnarray}
from which we can write that 
\begin{equation}
1={\frac{{\mu ^{2}\,e^{-2\,\psi }}}{{\mu ^{2}\,e^{-4\,\psi }-\omega ^{2}}}}\,%
{u^{0}}^{2}\,[1-v^{2}],
\end{equation}
where $u^{0}={\frac{{dt}}{{d\tau }}}$ is the usual time component of the
four velocity, and a definition of the spatial velocity, $v^{2}$, again
comes out naturally in this way. 
\begin{eqnarray}
v^{2} &=&{\frac{{e^{4\,\psi }\,(\mu ^{2}\,e^{-4\,\psi }-\omega ^{2})^{2}}}{{%
\mu ^{2}}}}\left( {\frac{{d\varphi }}{{dt}}}-{\frac{{\omega }}{{\mu
^{2}\,e^{-4\,\psi }-\omega ^{2}}}}\right) ^{2}+  \nonumber \\
&&+{\frac{{e^{2\,\gamma }\,(\mu ^{2}\,e^{-4\,\psi }-\omega ^{2})}}{{\mu ^{2}}%
}}\left( \left( {\frac{{d\rho }}{{dt}}}\right) ^{2}+\left( {\frac{{dz}}{{dt}}%
}\right) ^{2}\right),
\end{eqnarray}
which is the 3-velocity of a particle measured with respect to an
orthonormal reference system, it has components: 
\begin{equation}
v^{2}={v^{(\varphi )}}^{2}+{v^{(\rho )}}^{2}+{v^{(z)}}^{2}.
\end{equation}

For the $\varphi -$component of the spatial velocity we obtain: 
\begin{equation}
v^{(\varphi )}={\frac{{e^{2\,\psi }}}{{\mu }}}[(\mu ^{2}\,e^{-4\,\psi
}-\omega ^{2})\,\Omega -\omega ],  \label{Aeq:vphi}
\end{equation}
and substituting $\Omega $ from Eq.(\ref{Aeq:om2}), we finally obtain an
expression for the tangential velocity of a test particle in stable circular
motion: 
\begin{equation}
v^{(\varphi )}={\frac{{\mu \,e^{-2\,\psi }\,\sqrt{{\cal {A-B}}}-\omega \,%
\sqrt{{\cal {A}}}}}{{\mu \,e^{-2\,\psi }\,\sqrt{{\cal {A}}}-\omega \,\sqrt{%
{\cal {A-B}}}}}},  \label{Aeq:vphig}
\end{equation}
where ${\cal {A}}$ and ${\cal {B}}$ are given by Eqs.(\ref{Aeq:AB}).\newline

Imposing the condition of constancy for all radii, that is $v_{\rho
}^{(\varphi )}=0$, thus $v^{(\varphi )}=v_{c}^{(\varphi )}$, with $%
v_{c}^{(\varphi )}$ a constant, representing the value of the velocity, from
Eq. (\ref{Aeq:vphig}), we finally have that: 
\begin{equation}
{\cal {B}}=(1-{v_{c}^{(\varphi )}}^{2})\,F^{2}\,{\cal {A}},  \label{eq:cona}
\end{equation}
where $F=\left( {\mu }^{2}{\,e^{-4\,\psi }-\omega }^{2}\right) {/\left( {\mu
\,e^{-2\,\psi }+v_{c}^{(\varphi )}\,\omega }\right) }^{2}$. This last
expression, represents a constraint among three of the metric coefficients,
and we can express one of them, say $\omega $, in terms of the other two: $%
\psi $, and $\mu $. In this way, we again arrive to an iff condition,
namely: The tangential velocity of a test particle moving in a circular
equatorial motion in an axisymmetric stationary background, has a radii
independent magnitude iff the metric coefficients satisfy the constraint
equation (\ref{eq:cona}).

\end{document}